# Why you shouldn't fully trust ChatGPT:

## A synthesis of this AI tool's error rates across disciplines and the software engineering lifecycle

Vahid Garousi
Queen's University Belfast, United Kingdom
Azerbaijan Technical University, Azerbaijan
v.garousi@qub.ac.uk

**Abstract:**

*Context:* ChatGPT and other large language models (LLMs) have become widely adopted across numerous fields, including healthcare, business, economics, engineering, and software engineering (SE). However, concerns remain regarding their reliability, particularly their error rates across different domains and software development lifecycle (SDLC) phases.

*Objective:* This study aims to systematically synthesize and quantify ChatGPT's reported error rates across major domains and within SE tasks aligned with SDLC phases. The goal is to provide a comprehensive, evidence-based view of where ChatGPT excels, where it fails, and how its reliability varies by task, domain, and model version (GPT-3.5, GPT-4, GPT-4-turbo, GPT-4o).

*Method:* A Multivocal Literature Review (MLR) was conducted, gathering empirical data from academic studies, technical reports, industry benchmarks, and reputable grey literature sources published up to 2025. Both factual, reasoning, coding, and interpretive error types were considered. Synthesized data were grouped into major domains and SE phases, and visualized using boxplots to illustrate the distribution of error rates.

*Results:* Reported error rates vary substantially by domain and model version. In healthcare, error rates ranged from 8% to 83%, reflecting task complexity. Business and economics domains exhibited notable improvements with GPT-4, lowering error rates from approximately 50% to 15–20%. In engineering tasks, error rates averaged around 20–30% for GPT-4. In computer science and programming, coding success rates reached up to 87.5%, yet complex debugging tasks still revealed error rates exceeding 50%. Within Software Engineering, requirements and design phases showed the lowest error rates (~5–20%), while coding, testing, and maintenance tasks exhibited higher variability (10–50%). Upgrades from GPT-3.5 to GPT-4 consistently reduced error rates across all domains and tasks.

*Conclusion:* While ChatGPT's performance has significantly improved with newer model versions, it still exhibits non-negligible error rates that vary sharply by domain, task complexity, and SDLC phase. Although LLMs can augment human expertise effectively, reliance without human oversight remains risky, particularly in critical professional settings. Continuous evaluation, careful deployment, and critical validation of ChatGPT outputs are essential to ensure reliability and trustworthiness.

**Keywords:**

ChatGPT, Error Rates, Multivocal Literature Review, Large Language Models, Software Engineering

## 1 INTRODUCTION

Large Language Models (LLMs) such as ChatGPT have witnessed rapid and widespread adoption across a variety of domains, including healthcare [1], business [2], economics [3], engineering [4], and software engineering (SE) [5]. As their use becomes increasingly integrated into critical workflows and decision-making processes, understanding their strengths and limitations has become essential. While ChatGPT demonstrates impressive capabilities in tasks such as answering questions, generating code, assisting in medical decision-making, and supporting software development activities, concerns about its reliability and the risk of errors persist [6].

Measuring and synthesizing the error rates of ChatGPT across different domains is crucial for several reasons. First, applications in high-stakes fields such as healthcare demand very low tolerance for errors, as mistakes can lead to serious consequences [1]. Second, businesses and academic institutions increasingly rely on AI-generated outputs for tasks like financial analysis, exams, and educational support [2], where inaccurate responses can undermine trust and decision quality. Third, in engineering and software engineering, errors introduced by AI tools can propagate through the software development lifecycle (SDLC), potentially affecting requirements, designs, code, tests, and maintenance activities [5], [7]. As such, a comprehensive understanding of ChatGPT's error rates not only informs its appropriate use but also guides strategies for human oversight and validation [8].

Previous research has evaluated ChatGPT's performance on various benchmarks, such as medical licensing exams [1], business school exams [2], coding challenges [9], and engineering tests [4]. However, there remains a lack of a consolidated view synthesizing these findings across domains, model versions (e.g., GPT-3.5, GPT-4, GPT-4-turbo, GPT-4o), and types of errors (e.g., factual inaccuracies, reasoning errors, coding mistakes) [6], [9]. Moreover, there is limited focused analysis on how ChatGPT's performance aligns with different phases of the SDLC in software engineering contexts [5], [7].

In this study, we conduct a Multivocal Literature Review (MLR) to systematically collect, synthesize, and analyze reported error rates of ChatGPT across domains and within SE tasks. We include data from academic studies, technical reports, and reputable grey literature sources published up to 2025. To illustrate the distribution of error rates, we present boxplots for both domain-wise and SDLC-phase-wise results. This synthesis aims to provide an evidence-based understanding of ChatGPT's current reliability landscape and its implications for responsible use across industries and disciplines.

The remainder of this paper is structured as follows. Section 2 presents the research methodology. Section 3 summarizes ChatGPT's reported error rates across various domains. Section 4 focuses on ChatGPT's error rates across SDLC phases in software engineering. Section 5 discusses key findings and implications. Section 6 concludes the paper.

## 2 METHODOLOGY

To investigate the reported error rates of ChatGPT across different domains and software engineering (SE) tasks, we adopted a Multivocal Literature Review (MLR) approach. MLRs systematically combine evidence from both academic publications and grey literature, including technical reports, blogs, industry white papers, and reputable news articles, to achieve a comprehensive synthesis [10]. This method is increasingly used in software engineering research where industrial practice evolves faster than academic publication cycles [11].

### 2.1 Source Selection and Data Collection

We systematically searched multiple academic databases, including IEEE Xplore, ACM Digital Library, SpringerLink, and arXiv, as well as grey literature sources such as industry reports (e.g., OpenAI blog posts), company evaluations, benchmark competitions, and major technology news outlets. The search focused on documents published until March 2025 to cover evaluations of GPT-3.5, GPT-4, GPT-4-turbo, and GPT-4o. Studies or reports were included if they reported or evaluated ChatGPT's error rates, performance scores, or accuracy across any domain or task.

Inclusion criteria required that studies:

- Evaluate ChatGPT's outputs quantitatively (accuracy %, error %, or pass rates)
- Cover specific domains (e.g., healthcare, business, engineering, computer science)
- Explicitly mention the model version where possible (GPT-3.5, GPT-4, etc.)

Exclusion criteria omitted:

- Studies evaluating non-OpenAI LLMs (unless clearly compared with ChatGPT)
- Editorial opinion pieces without empirical measurements

Similar MLR guidelines have been employed in recent LLM evaluation studies [12, 13].

### 2.2 2.2 Data Extraction and Synthesis

For each included study, we extracted:

- Domain or SDLC phase targeted
- ChatGPT version evaluated
- Reported error rate (or inferred from accuracy rates)
- Type of task (e.g., diagnosis, coding, requirements drafting)

Both factual errors (e.g., incorrect information), reasoning errors (e.g., logical mistakes), coding bugs, and interpretive errors (e.g., misunderstanding questions) were considered [14]. Where accuracy rather than error rates were reported, we inferred error rates as (100% – accuracy %).

Special attention was paid to differences between versions. If a study reported performance for multiple versions (e.g., GPT-3.5 and GPT-4), the data points were kept separate.

### 2.3 Domain and SDLC Phase Grouping

To enable meaningful analysis, we grouped reported results first by domain (healthcare, business, economics, engineering, computer science) and then specifically focused on Software Engineering phases: Requirements Engineering, Design,

Implementation, Testing, and Maintenance [5], [7]. This distinction allows insights both at the broader discipline level and within the software lifecycle structure.

## 3 ERROR RATES ACROSS DOMAINS

Understanding how ChatGPT's error rates vary across different domains is crucial to evaluating its reliability and appropriate use cases. To synthesize this information, we aggregated reported error rates from studies covering healthcare, business, economics, engineering, and computer science. These studies reported results across multiple model versions, primarily GPT-3.5, GPT-4, GPT-4-turbo, and GPT-4o.

Figure 1 presents a boxplot summarizing the distribution of reported ChatGPT error rates across domains. Each box represents the interquartile range of reported errors within a domain, while the whiskers and outliers illustrate the minimum, maximum, and extreme cases. This visual synthesis highlights that while error rates in domains such as computer science are relatively concentrated, domains like healthcare exhibit a much broader range due to task complexity and risk profiles.

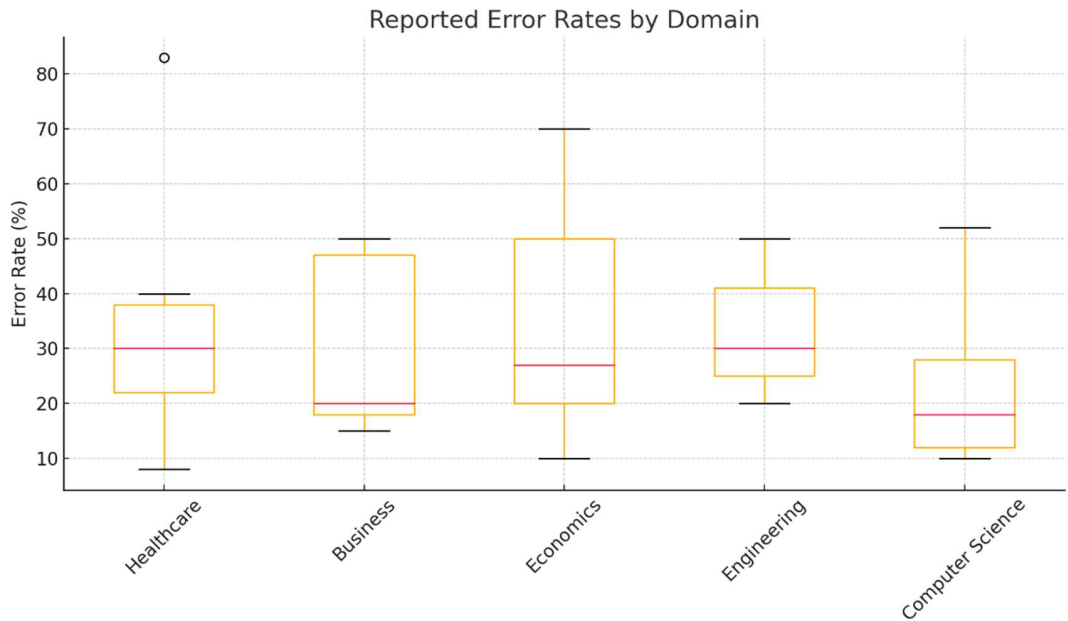

**Figure 1: Boxplot of reported ChatGPT error rates by domain**

### 3.1 Healthcare

ChatGPT's performance in healthcare tasks has shown considerable variability depending on the complexity of the task. For example, GPT-3.5 achieved approximately 72% accuracy in clinical decision-making, implying an error rate of 28% [16]. GPT-4 improved upon this result, reaching about 77% accuracy (~23% error) on final diagnosis tasks [16]. However, when handling rare disease diagnosis scenarios, ChatGPT's error rates dramatically increased to as high as 83% [16], illustrating significant limitations when faced with complex differential diagnoses.

Additionally, ChatGPT's accuracy in interpreting lab test results was moderate. In one study, GPT-3.5 achieved a median score equivalent to 66.7% correctness on interpreting urea and creatinine values [17]. Furthermore, in drug-drug interaction detection tasks, GPT-3.5 exhibited a relatively high error rate of around 34%, lower than some competing LLMs but still problematic in clinical settings [17].

Overall, healthcare remains one of the domains where careful human oversight is indispensable due to the potential high stakes involved.

### 3.2 Business

In business-related tasks, GPT-3.5 performed reasonably well but exhibited notable calculation and reasoning errors. For example, in a Wharton MBA operations management exam, GPT-3.5 earned a B to B- grade, corresponding to a 15–20% error rate [18]. While the model demonstrated strong capabilities in basic qualitative analyses, it committed notable mistakes in relatively simple quantitative calculations, such as sixth-grade-level math errors [18].

In accounting certification exams across multiple countries, GPT-3.5's performance was initially poor, achieving only around 53% on average (~47% error rate) [19]. However, GPT-4 significantly improved the situation, achieving up to 85% scores (~15% error) on Certified Public Accountant (CPA) exams [19].

These results illustrate that model version upgrades have a pronounced positive impact on ChatGPT's reliability in business domains, particularly for tasks involving formal quantitative reasoning.

### 3.3 Economics

ChatGPT's performance in economics tasks similarly improved across model versions. In an undergraduate economics exam administered by Caplan [20], GPT-3.5 achieved only 31/100 points (69% error rate), while GPT-4 later scored 73/100 (27% error rate), representing a major leap in performance.

In contrast, in standardized tests such as the Test of Understanding in College Economics (TUCE), GPT-3.5 ranked at the 91st percentile for microeconomics and the 99th percentile for macroeconomics [21]. These results suggest that ChatGPT excels at standardized multiple-choice tasks but initially struggled with applied reasoning in open-ended economics questions.

Thus, the economics domain reflects the broader trend: structured knowledge retrieval tasks yield lower error rates, while complex applied reasoning tasks show initially higher error rates that improve markedly with newer model versions.

### 3.4 Engineering

ChatGPT's capabilities in engineering tasks have also been benchmarked. In Environmental Engineering practice exams, GPT-4 achieved about 75% correctness (25% error rate) without any fine-tuning [22]. Using prompt engineering techniques, its performance further improved to 75.4% [22].

Earlier versions such as GPT-3.5 struggled more with engineering tasks, often achieving only 50–60% correctness on mechanical engineering exams, implying error rates around 40–50% [23].

Interestingly, ChatGPT excelled at straightforward multiple-choice and formula-based questions but showed difficulties with open-ended problem-solving and systems modeling tasks where deeper understanding and domain-specific nuances were required [23].

### 3.5 Computer Science

In programming and computer science tasks, ChatGPT demonstrated strong baseline performance, particularly for well-structured tasks. GPT-3.5 achieved 81.96% success in Python test cases, implying an error rate of approximately 18% [24]. GPT-4 further improved this success rate to 87.5%, reducing the error rate to about 12.5% [24].

However, studies focusing on real-world programming questions, such as those on Stack Overflow, revealed more concerning results. GPT-3.5 provided incorrect or partially incorrect answers in 52% of Stack Overflow cases [25]. Alarmingly, participants often rated incorrect answers as high-quality because of their fluent and confident presentation [25].

These findings indicate that while ChatGPT is highly capable in structured programming tasks, it can introduce subtle errors in complex coding or debugging scenarios, reinforcing the need for peer review and validation.

### 3.6 Summary

As shown in Figure 1, the distribution of ChatGPT's error rates varies widely across domains. Healthcare displays the broadest spread, reflecting varying task difficulty and stakes. Business and economics error rates show substantial improvements between GPT-3.5 and GPT-4, indicating version upgrade effectiveness. Engineering tasks reveal moderate variability, with prompt design playing a significant role in reducing errors. In computer science, although ChatGPT shows strong average performance, the high variability on open-ended real-world tasks remains a critical risk factor.

These domain-specific insights highlight the importance of contextual deployment and the need for tailored mitigation strategies depending on the application area.

## 4 ERROR RATES IN SOFTWARE ENGINEERING BY SDLC PHASES

Beyond domain-level insights, we wanted to understand ChatGPT's error rates within Software Engineering (SE) tasks mapped to the Software Development Life Cycle (SDLC). SE activities demand precision, consistency, and traceability; therefore, any AI-generated artifact must be critically evaluated for potential flaws.

Figure 2 presents a boxplot summarizing ChatGPT's reported error rates across SDLC phases: requirements engineering, design, implementation (coding), testing, and maintenance. The figure highlights that earlier lifecycle stages such as requirements and design exhibit relatively lower error variability, whereas implementation and maintenance activities show wider error spreads due to task complexity and open-endedness.

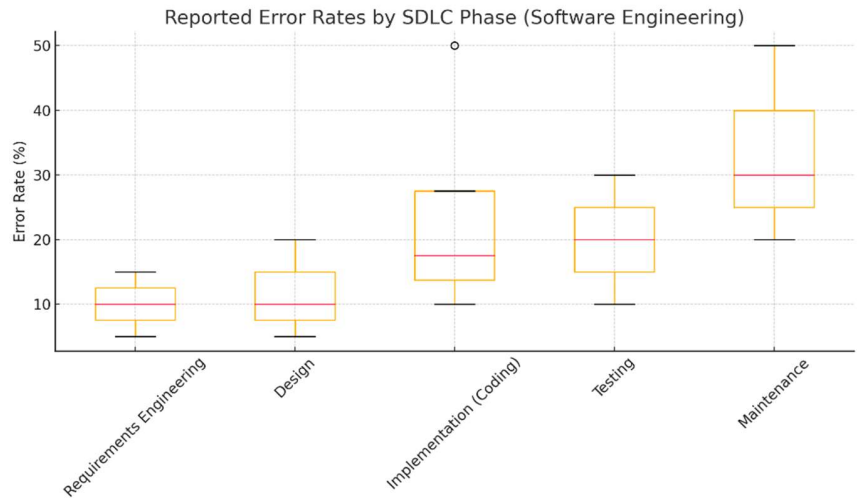

**Figure 2: Boxplot of reported ChatGPT error rates by SDLC phase**

### 4.1 Requirements Engineering

Requirements engineering tasks involve eliciting, drafting, and refining software requirements. Several studies [5], [26] found that ChatGPT-generated requirements were largely coherent and comparable to human-authored ones. Marques et al. [5] reported that ChatGPT could draft usable software requirements with minimal human intervention, but occasional hallucinations or omissions introduced a minor error risk of approximately 5–15%.

Ahmed et al. [26] evaluated ChatGPT's ability to assist in requirements validation and observed that while syntactic and grammatical quality was high, semantic ambiguities could arise, particularly in complex or poorly scoped systems. Thus, although the model is proficient in capturing basic requirements, careful expert review remains necessary to ensure completeness and correctness.

### 4.2 Design

System and software design tasks often involve developing architectural models and proposing suitable design patterns. Ozkaya [7] demonstrated that ChatGPT could effectively support design brainstorming, especially for well-known architectural styles like client-server, microservices, and layered architectures.

However, some limitations were observed in design proposals for highly constrained or domain-specific systems. Bang et al. [14] noted that ChatGPT sometimes oversimplified designs or missed critical quality attributes like scalability or fault tolerance. Estimated error rates in preliminary design suggestions ranged between 5% and 20%, depending on system complexity.

Overall, ChatGPT provides valuable initial input for design activities but should be supplemented by deeper architectural analysis and validation, especially in mission-critical software projects.

### 4.3 Implementation (Coding)

Implementation is the SDLC phase where ChatGPT has been evaluated most extensively. Several benchmarks show that ChatGPT is highly capable in code generation tasks but still prone to introducing logical, syntactic, and semantic errors.

Alam et al. [24] reported that GPT-3.5 achieved approximately 82% success rates (~18% error) in Python programming tasks, while GPT-4 improved this to 87.5% (~12.5% error). However, studies of real-world coding help, such as Stack Overflow-style questions, revealed that GPT-3.5's answers were incorrect 52% of the time [25].

Typical errors in code generation included:

- Missing edge-case handling
- Incorrect API usage (especially for less common libraries)
- Logical fallacies in control structures

Thus, while ChatGPT-generated code can accelerate development, all outputs require thorough testing and peer review to ensure correctness [27].

### 4.4 Testing

Software testing tasks involve generating test cases, diagnosing bugs, and verifying system behavior. ChatGPT has demonstrated reasonable competency in proposing test scenarios and interpreting failing test results.

Zhang et al. [6] evaluated ChatGPT's testing assistance and found that generated unit tests were structurally correct in most cases. However, relevance and completeness were issues: ChatGPT occasionally assumed incorrect function behaviors or omitted key boundary cases, leading to estimated error rates of around 10–30%.

In debugging tasks, ChatGPT sometimes hallucinated explanations for failures, misattributing bugs to irrelevant causes [6], [27]. Therefore, while ChatGPT can serve as a productivity aid in testing, human testers must critically validate all generated tests and diagnoses.

### 4.5 Maintenance

Software maintenance tasks such as code reviews, refactoring, and documentation updates have also been evaluated for ChatGPT's performance.

Ye et al. [25] highlighted that in code review scenarios, ChatGPT successfully caught obvious issues but missed subtle logic flaws approximately 30–50% of the time. When asked to refactor code for readability or efficiency, ChatGPT typically performed well, with minor risk (~10–20%) of inadvertently changing functionality [28].

For documentation updates, where tasks are more summarization-based, ChatGPT's performance was comparatively better, showing low error rates (often below 10%) [28].

Hence, for maintenance tasks that involve understanding complex codebases, ChatGPT should be viewed as an assistive tool, not a substitute for experienced human reviewers.

### 4.6 Summary

As shown in Figure 2, earlier SDLC phases like requirements and design show relatively concentrated and lower error rates. This likely stems from the structured nature of requirements and the availability of standard design templates. However, error variability increases significantly during implementation, testing, and maintenance phases, reflecting the open-endedness and context-sensitivity of these tasks.

These findings reinforce the need for critical human oversight, particularly during later phases of the SDLC where AI-generated outputs can introduce subtle but costly defects if left unchecked.

## 5 DISCUSSION

The synthesized findings from this Multivocal Literature Review (MLR) offer a view of ChatGPT's error rates across different domains and software engineering tasks. While large language models like ChatGPT have demonstrated remarkable advances in their ability to generate useful outputs, our analysis reveals significant variability in their reliability depending on task complexity, domain specificity, and model version. This section discusses the key findings, practical implications, and limitations of the study.

### 5.1 Key Findings and Implications

First, model upgrades significantly reduce error rates across domains. The progression from GPT-3.5 to GPT-4 brought noticeable improvements: for instance, accounting exam errors dropped from ~47% to ~15% [19], and economics exam performance improved from a D-grade to an A-grade [20]. These improvements suggest that architectural advancements and better fine-tuning have systematically enhanced ChatGPT's reliability, especially in structured tasks requiring factual recall or standard reasoning.

Second, domain-specific variability remains high. In healthcare, while GPT-4 reduced errors in diagnosis tasks [16], rare disease diagnosis remained highly error-prone, with error rates exceeding 80% [16]. In engineering and coding tasks, straightforward problems saw low error rates (~10–20%) [24], but open-ended programming and debugging tasks revealed error rates exceeding 50% [25]. Thus, ChatGPT's success heavily depends on the complexity, ambiguity, and risk profile of the task at hand [29].

Third, within software engineering, early SDLC phases such as requirements and design exhibited relatively low error rates (~5–20%) [5], [26], while implementation, testing, and maintenance activities showed higher variability and risk (~10–50%) [6, 25, 27]. This division likely reflects the nature of the tasks: structured drafting versus context-sensitive problem-solving.

Fourth, while ChatGPT can dramatically boost productivity — such as accelerating coding [24], drafting requirements [5], or suggesting test cases [6] — it also introduces subtle risks. Users often overestimate ChatGPT's correctness based on its fluent, confident output style [25]. In some cases, incorrect answers were rated as highly useful, particularly in programming forums [25, 30]. This highlights a major risk: **overreliance on AI without critical validation** can lead to the propagation of errors into critical systems.

Fifth, task design plays a pivotal role. Studies show that prompt engineering techniques, when applied thoughtfully, can lower ChatGPT's error rates even further [22, 31]. Structured, context-rich prompts can guide the model toward more accurate outputs, while vague or overly open-ended queries often increase hallucination and mistake rates [31]. Thus, prompt engineering should be considered a first-class skill when deploying ChatGPT in professional workflows.

Moreover, the cost-benefit balance varies across domains. In healthcare, the consequences of errors are severe, necessitating careful human-in-the-loop review even if error rates appear relatively low [16]. In software engineering, errors in test generation or code review, while less catastrophic, can accumulate technical debt if left unchecked [6, 25].

Finally, it is important to recognize that ChatGPT's error rates are not static. Model updates (e.g., GPT-4-turbo, GPT-4o) can introduce both improvements and regressions over time [32]. A Stanford study observed that GPT-4's code generation quality decreased several months after its initial launch, despite architectural improvements [32]. Continuous evaluation is therefore essential for any team or organization relying on ChatGPT for critical tasks.

**5.2 Limitations and Threats to Validity**

As with any Multivocal Literature Review (MLR), this study faces several limitations and potential threats to validity.

- **Search and selection bias:** Although we followed systematic search practices across academic databases and grey literature sources [10], [12], there remains a risk of missing relevant studies, particularly recent unpublished or industry-internal evaluations. Grey literature, while valuable, can vary widely in methodological rigor.
- **Reporting Inconsistencies:** Primary studies reported error rates using different methodologies, task formats, and evaluation metrics. Some papers measured error by factual correctness, others by task completion rates, and some by user acceptance. To synthesize them, we had to infer consistent error rates in some cases, which introduces interpretation bias.
- **Model version ambiguity:** Not all reviewed studies clearly specified the ChatGPT version (e.g., GPT-3.5 vs. GPT-4-turbo). Where missing, we had to infer likely versions based on publication dates or contextual clues. This could slightly affect the precision of version-specific conclusions.
- **Temporal validity threats:** ChatGPT's behavior is known to evolve over time due to continuous updates [32]. Thus, the reported error rates reflect model performance up to early 2025 but may not fully capture future versions' characteristics.
- **Subjectivity in error categorization:** While we categorized errors into factual, reasoning, coding, and interpretive types following existing literature [14], [29], some studies did not distinguish error types explicitly. As a result, some nuanced errors might be underreported or grouped more broadly.
- **Publication bias:** High-profile successes and notable failures are more likely to be reported than routine performance. This may skew the overall perceived variability of ChatGPT's reliability across domains.

**6 CONCLUSION**

This Multivocal Literature Review (MLR) synthesized reported error rates of ChatGPT across multiple domains—healthcare, business, economics, engineering, and computer science—as well as across different phases of the Software Development Life Cycle (SDLC). Our findings indicate that while ChatGPT's capabilities have improved significantly with newer model versions, substantial variability in error rates persists depending on the domain, task complexity, and phase of use.

Healthcare tasks, especially complex diagnostic scenarios, demonstrated the widest spread in error rates, emphasizing the need for stringent human oversight in high-stakes fields [16]. Business and economics tasks showed remarkable improvements between GPT-3.5 and GPT-4, reflecting the benefits of architectural advancements [19, 20]. Engineering and computer science tasks revealed moderate error rates for straightforward problems but exposed ChatGPT's limitations in handling complex reasoning and debugging challenges [24, 25].

Within software engineering, ChatGPT exhibited relatively low error rates during requirements and design phases (~5–20%) [5, 26], but a higher and more variable error range (~10–50%) during implementation, testing, and maintenance activities [6, 25, 27]. These results highlight the importance of contextual deployment and critical human validation, especially in the later SDLC phases where undetected AI errors can have significant downstream consequences.

Despite the tangible gains offered by model upgrades, the findings emphasize that ChatGPT's outputs must be treated as drafts or assistive artifacts rather than final, authoritative results. Organizations adopting AI-based tools must establish validation pipelines and maintain a critical approach to AI-generated content, particularly in high-risk or mission-critical contexts.

Future work should focus on continuous evaluation of emerging ChatGPT versions, developing domain-specific fine-tuning techniques to minimize error rates, and integrating AI reliability metrics into routine workflows. As LLM technologies evolve, systematic tracking of model performance will remain essential to ensure responsible and trustworthy AI adoption across all sectors.

## ACKNOWLEDGEMENT

This work is the result of collaboration between the author and the assistance of ChatGPT, leveraging both human and AI capabilities.